# LSTM Based Music Generation System

Sanidhya Mangal
Computer Science and Engineering
*Medi-Caps University*
Indore, India
mangalsanidhya19@gmail.com

Rahul Modak
Computer Science and Engineering
*Medi-Caps University*
Indore, India
rahulsvmodak@gmail.com

Poorva Joshi
Computer Science and Engineering
*Medi-Caps University*
Indore, India
purvaj27@gmail.com

*Abstract*— Traditionally, music was treated as an analogue signal and was generated manually. In recent years, music is conspicuous to technology which can generate a suite of music automatically without any human intervention. To accomplish this task, we need to overcome some technical challenges which are discussed descriptively in this paper. A brief introduction about music and its components is provided in the paper along with the citation and analysis of related work accomplished by different authors in this domain. Main objective of this paper is to propose an algorithm which can be used to generate musical notes using Recurrent Neural Networks (RNN), principally Long Short-Term Memory (LSTM) networks. A model is designed to execute this algorithm where data is represented with the help of musical instrument digital interface (MIDI) file format for easier access and better understanding. Preprocessing of data before feeding it into the model, revealing methods to read, process and prepare MIDI files for input are also discussed. The model used in this paper is used to learn the sequences of polyphonic musical notes over a single-layered LSTM network. The model must have the potential to recall past details of a musical sequence and its structure for better learning. Description of layered architecture used in LSTM model and its intertwining connections to develop a neural network is presented in this work. This paper imparts a peek view of distributions of weights and biases in every layer of the model along with a precise representation of losses and accuracy at each step and batches. When the model was thoroughly analyzed, it produced stellar results in composing new melodies.

## I. INTRODUCTION

The art of ordering tones or sound in succession, in combination is music. It is a temporal relationship to produce a composition of notes having continuity and unity. In other words, a musical note is any sound generated with the help of musical instruments or human voice. A musical note is a simple unit of music. Music and its notes have certain properties [1] concerning its quality and performance.

The sound input for training the Artificial Intelligence (AI) model can be monodic, having a single melodic line, or polyphonic, involving many sounds. The musical notes are exhibited by an octave or some interval of pitches. Pitch has a pitch class that refers to the relative position in an octave. Music details are essential for training the model and the model's complexity and output depends upon the nature of the input. The model to be trained is supposed to recall the former details and create a rational piece. Music is generated by playing notes of different frequencies and the linkages between the notes are preserved.

One method of generating music by utilizing existing music is genetic algorithm [2]. As stated in [2] genetic algorithm can highlight the strong rhythm in each fragment and combine them into distinct pieces of music. But it has low efficiency because every iteration process of it has a delay. In addition, due to the lack of context, it is difficult to get the coherence and deep-seated rhythm information.

So, to overcome the above problem we require a system that should be able to remember the previous note sequence and predict the next sequence and so on. Recurrent Neural Networks [3] especially Long Short-term Memory a special RNN is used.

This paper construes an algorithm (Neural Network) based on LSTM networks which can be used to generate music and melodies automatically without any human intervention. The key goal is to develop a model which can learn from a set of musical notes, analyze them and then generate a pristine set of musical notes. This task is a real challenge because the model must have capabilities to recall past details and structure of musical notes for future projection of learning sequence. The model needs to learn the original sequences adjacent to past one and transform it for the learning system.

Additionally, this paper covers technical challenges, software implementation of proposed model and related work pursued by different researchers in this domain. In context to performance and efficiency, their work is cited and analyzed to produce stellar results for this model. In later sections of this paper, a closer analysis of model's performance and efficiency is discussed in detail along with outliers and proper plots of curve generated by training and validating this model.

This paper is organized in the following manner: Section II provides a glimpse of related work done in the field. Section III introduces the theoretical concept involved along with the technical challenges faced. The proper implementation of an algorithm describing designs of deep neural network and software is presented in Section IV. Section V presents the result simulation of developed LSTM model for music generation and a brief comparison with past work is also discussed. Conclusions, Acknowledgement and References are in Section VI.

## II. RELATED WORK

Music generation is a topic that has been studied in much detail in the research industry in the past. There are several who tried to generate music with different approaches. Garvit et al. [4] also presented a paper to compare these algorithms on two different hardware's.

There are several approaches which intend to generate a suite of music and their combination can be used to design a new and efficient model. We divide these approaches into two wide categories. One is traditional algorithms operating on predefined functions to produce music and another is an autonomous model which learns from the past structure of musical notes and then produces a new music.

Drewes et al. [5] proposed a method on how can algebra be used to generate music in a grammatical manner with the help of tree-based fashion. Markov chains [6] and Markov hidden units can be used to design a mathematical model to generate music.

After the breakthrough in AI, many new models and methods were proposed in the field of music generation. Description of various AI enabled techniques can be found in [7], [8], [9] including a probabilistic model using RNNs, Anticipation-RNN and recursive artificial neural networks (RANN) an evolved version of artificial neural networks [10] for generating the subsequent note, subsequent note duration, rhythm generation. Generative adversarial networks (GANs) [11] are actively used in generating musical notes which contain two neural networks, generator network that generates some random data and discriminator network that evaluates generated random data for authenticity against the original data(dataset). MuseGAN[12] is a generative adversarial network that generates symbolic multi-track music.

There exist many open source systems to generate music, e.g. [13], but an in-depth discussion of all such research works and methods is beyond the scope of this document.

## III. METHOLODGY

This model is inferred from "Generation of Music using LSTM" [14]. In their paper, a Biaxial LSTM model i.e. two LSTMs are used. One to predict the time at which the node should be played and another to predict that note. Here we are using one LSTM to predict both, the time and the note in sequence.

### A. Technical Challenges

One key obstacle is the representation of data. Here, we selected MIDI file format because firstly, it bears characteristics of a song in its metadata and secondly, it is commonly used, as considerable number of datasets are available.

Recurrent neural networks experience a serious problem known as vanishing or exploding gradient problem. The problem is that they might not be able to connect information from several previous steps to the present step. This phenomenon was explored by Hochreiter (1991) [15] and Bengio, et al. (1994) [16]. They proposed a solution called LSTM (Long short-term memory) cells which are capable of remembering long-term dependencies which are used in this paper.

## IV. IMPLEMENTATION

This section contains details about the model which went under the process of training and validation. A fully trained model was used to generate a suite of music. Experiment and model training were conducted on Google Colab, running over Google Cloud Platform with deep learning and code implementation on Keras (using Tensorflow as backend).

### A. Deep Neural Network Design

The model is applied on polyphonic musical notes. The LSTM network is trained to acquire the knowledge of probability of occurrence of a musical note at current time. The output of the network at a time step t, conditioned on the previous notes' state till time step t-50 are fed into the input unit to recall past details and structure of notes.

The LSTM layer depends on a selected input. Not all notes undergo the training process. Only some selected and specified notes are used to train this LSTM model, which are useful for effectively tuning the model, that result in efficient information gain. With these inputs, LSTM layer learns the mapping and correlation between notes and their projection. Next to the LSTM layer, Dropout [17] layer is used to create generalizations in the model. Once the model has learned the probability distributions of notes and sequences, we must combine all the LSTM cells with each other. This gap is subordinated with the help of Dense [18] layer. Dense layer ensures that the model is fully connected. At the endpoint, the Activation layer is added to the model, which helps in deciding, which neurons (LSTM cells) should be activated and whether the information gained by the neuron is relevant, making activation function highly important in a deep neural network.

After training the LSTM network, the model is ready to generate a new sequence of musical notes. To ensure better prediction and diverse output of sequences, a large and varied dataset was elicited with different variations in the structural composition of musical notes. The goal was to expose the model with diverse dataset which would lead to a better tuning of the model. The MIDI file format was used to extract dataset. MIDI files played an important role in extracting information about note sequence, note velocity and the time component.

A model was compiled using Moon et al. [19] as a suggested guide for dropouts, Dropout of 0.75 was applied to the LSTM layer. The optimizer selected was RMSprop [20]. The learning rate selected was 1e-4 for model optimization.

### B. Software Design

Data is one of the important factors in training and validating a neural network. The data vector used in this work (neural network) is referred to "Note Matrix". Note, note velocity and the time interval of a note are the attributes that collectively form a "Note Matrix". "Note Matrix" is not directly fed into the model, it undergoes some preprocessing and filtering. This process can be divided into two stages. At the first stage, the data is cleaned and scaled between (0, 1) for better learning and performance. Next, the data is divided into 3D tensor i.e. "Note Matrix", contains an internal batch to determine the probabilistic occurrence of the next sequence. Mido module extracted from [21] is used to perform pre/post-processing of "Note Matrix", including the creation of MIDI file for generated notes by the model and importing MIDI files and tracks.

The overall blueprint of the code can be distributed into two major tasks: training/validating the model and generating a suite of music after model training. Same model and graph are used by both the functions. The training task shown in fig. 1 takes input iteratively as a "Note Matrix", runs the model and computes the probabilistic occurrence of notes along with the losses and accuracy at corresponding time step and for all the notes present in the batch. After successful training of the model, it predicts the set of note sequence at a given time step.

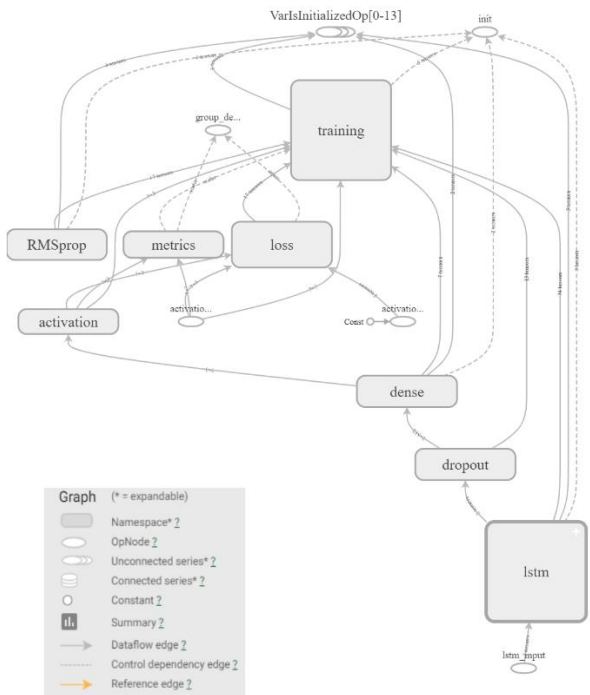

Fig. 1. This image provides an insights of data flow in the model. It describes complete component structure of the model used in training process. It also shows intermediate steps used in training process of the model along with the metrics, initializers and variables. The lines connecting components show number of tensors flowing through from one component to another.

The actual model as shown in fig. 2, can be broken down into five functional modules:1) LSTM layer composed of 512 neurons(units) which consists of three components: bias, kernel and recurrent kernel. Bias acts as a threshold for meaningful activation of neurons, the kernel takes an input of "Note Matrix" along with the output of recurrent kernel so that the model can recall the past details about the note structure and project them for future use. Fig. 4 shows distributions of weights in the LSTM layer. 2) Dropout layer, to generalize the training portion done by LSTM layer. 3) Dense layer, to connect all the LSTM neurons with each other and to produce the coveted output required by the user. In this case, it produces an output similar to "Note Matrix". 4) Activation layer, to determine which neuron should be active and to produce final output of same dimension as that of the layer above it. It also determines the probability of occurrence of note sequence. In this model, "Linear Activation" [22] is used. 5) RMSprop, an optimizer used to compile model and compute losses.

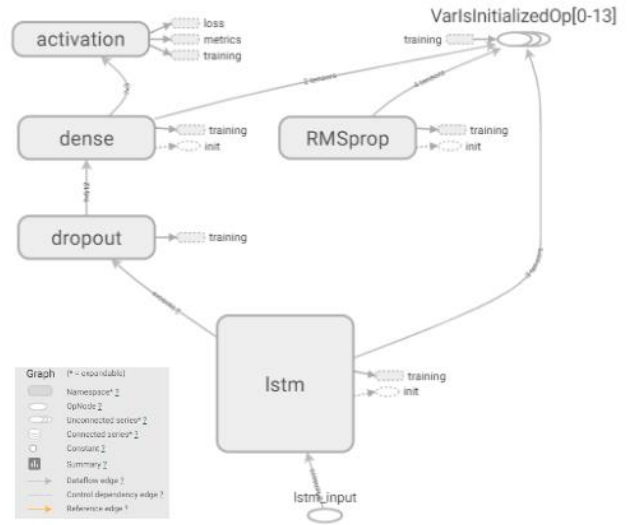

Fig. 2. This image provides the structure of graph used to implement this model, representing various components and classes used to create this model. The image is generated by Tensorboard.

The loss function used is mean squared error (also known as L2 loss), shown in fig. 3. It calculates the distance between actual and predicted sequence and then averages its sum over complete data set.

= the square of the difference between the label and the prediction
= (observation - prediction(x))²
= (y - y')²

**Mean square error (MSE)** is the average squared loss per example over the whole dataset. To calculate MSE, sum up all the squared losses for individual examples and then divide by the number of examples:

$$MSE = \frac{1}{N} \sum_{(x,y) \in D} (y - prediction(x))^2$$

where:
- $(x, y)$ is an example in which
  - $x$ is the set of features (for example, chirps/minute, age, gender) that the model uses to make predictions.
  - $y$ is the example's label (for example, temperature).
- $prediction(x)$ is a function of the weights and bias in combination with the set of features $x$.
- $D$ is a data set containing many labeled examples, which are $(x, y)$ pairs.
- $N$ is the number of examples in $D$.

Fig. 3. Shows pseudo code for MSE and formulae to compute MSE, taken from [23].

Link to the code of this work and output summary of Tensorboard produced after training this model is provided [24].

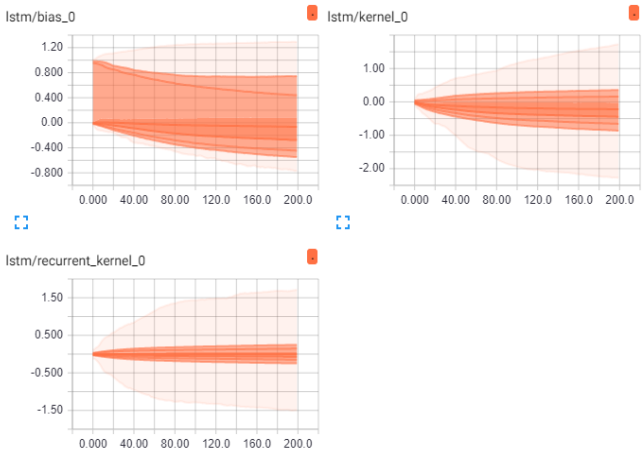

Fig. 4. Image describing distributions of weights in various stages of the LSTM layer used in this model. LSTM/bias depicts distribution of biases across the layer, LSTM/kernel describes weights in current input and time state in the model and LSTM/recurrent_kernel describes weights for previous state and time stored to provide ability to recall past details and structure of model. These two kernels and biases collectively forms an LSTM layer. Image is taken from Tensorboard.

## V. RESULTS

Fig. 5 shows the performance of past papers, as well as the results obtained by training this model. The result captured by this model is a slight improvement compared to previous work. Performance of the model was profoundly affected due to less exposure to training time. In their blog [14], they mentioned that several authors trained their model for 22-24 hours to tune it for better sequence capturing whereas, training this model took only 0.9 hours (54 mins) on Pop Music [25].

TABLE I. RESULT ANALYSIS OF DIFFERENT MODELS

| Model | Log Likelihood | Hours Trained |
|---|---|---|
| Random | -61 | - |
| TP-LSTM-NADE | -5.44 | 24-48 |
| BALSTM | -4.90, -5.00 | 24-48 |
| BALSTM | -5.16, -6.59 | 16 |
| Single LSTM (this work) | -6.23 | 0.9 |

Fig. 5. Top 4 rows represent the Log-likelihood of performance reported by Nikhil and Paul [14] for random weighted, TP-LSTM-NADE, Bi-axial LSTM (original author) and their model was tested on Piano-Midi.de dataset. The two values are regarded as the best and median performances across 5 trials. The tail row represents the best value across 200 trials on single layered LSTM network trained on [25] dataset.

More perspicacity concerning the training of model can be gained by looking at training graphs. It can be seen in the fig. 6 that loss gradually decreases as we move towards higher number of epochs(iterations). It is observed that at the starting point the loss was approximately 0.0286 but as we move further, loss function converges rapidly and then becomes steady and slow towards the end and gets reduced approximately to 0.002098 at 200th epoch. At 120th epoch there is a tremendous increase in the loss which is an outlier in this process. The loss shoots up to approximately 0.0718.

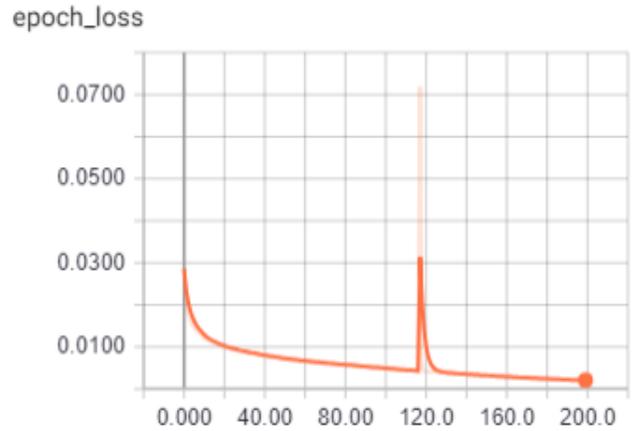

Fig. 6. Final training image of epoch (iterations) loss of our model used on Piano Midi files for 200 epochs.

Fig. 7 describes the accuracy of the model for all 200 epochs. It is observed that accuracy at initial level is 71%, it increases linearly and from 20th to 25th epoch, it reaches 90%. After 25th epoch, accuracy increases with smaller gradient and in steady manner. Despite of outliers at 120th step, accuracy doesn't get affected much, explicitly defining stability of this model. At the last epoch, accuracy projected by the model is 97.23%.

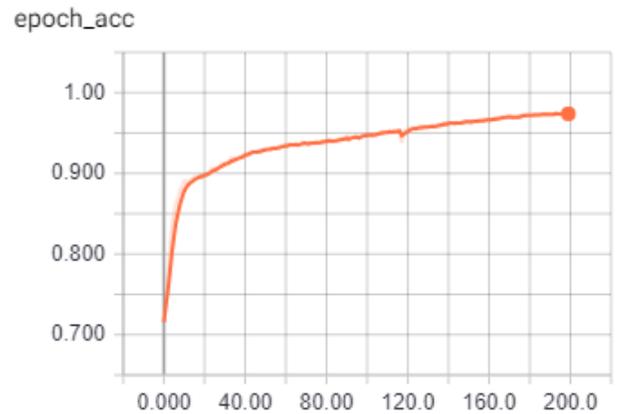

Fig. 7. A graph describing training accuracy of 200 epochs on Piano Midi files, this graph is a pair of fig..6.

The model takes an input of "Note Matrix" in the form of batch, which is used to tune weights and biases in the model. On taking a look on batch losses, shown in fig. 8 and batch accuracy in fig. 9, it is observed that the model got tweaked approximately 1,460,000 times.

There are jogs in the fig. 8 and fig. 9 which depict that new batches are sampled. On taking a closer look at fig. 7 and fig. 9, it is observed that there is not much change in accuracy projected by them but in case of loss projected by batch and by epoch, there is slightly more difference than expected. Value of loss projected in fig. 8 at the last epoch is 0.00196 and that in fig. 9 is 97.27%.

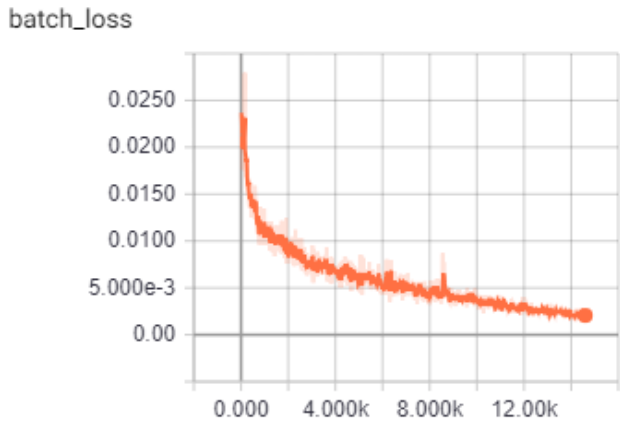

Fig. 8. Graph depicting training loss of batches in the model, generated by Tensorboard. See text for details.

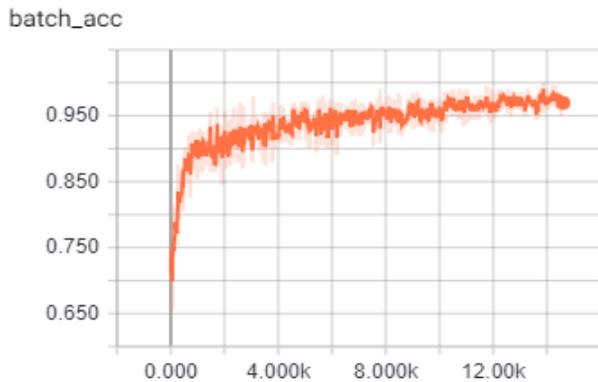

Fig. 9. Graph Depicts batch accuracy of the model used in this work, generated by Tensorboard. See text for details.

More details about the model can be gathered from TensorBoard visualizations. Distributions of data changes over time such as a weight of neural network and a fancier look of these distributions can be visualized through histograms by the 3D look obtained. Fig. 10 shows a small peek view of Distributions and fig. 11 shows a glimpse of Histograms produced by these distributions. Multiple authors reported Adadelta[26] for model optimization but this work still uses RMSprop.

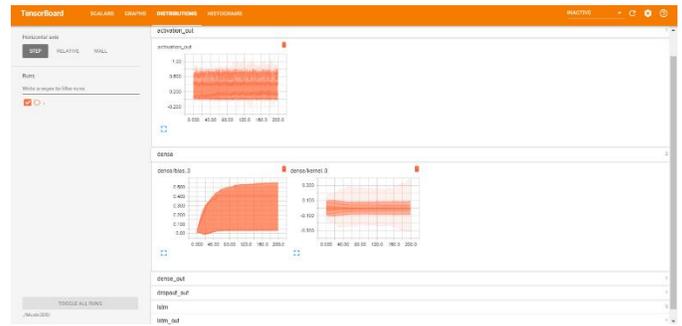

Fig. 10. Image showing weight distributions of the model. It shows an expanded view of output of activation layer and distributions of the dense layers, distributions of LSTM layer and Output form Dense, Dropout and LSTM layer can also be visualized by running Tensorboard server on model summary. Image generated by the Tensorboard.

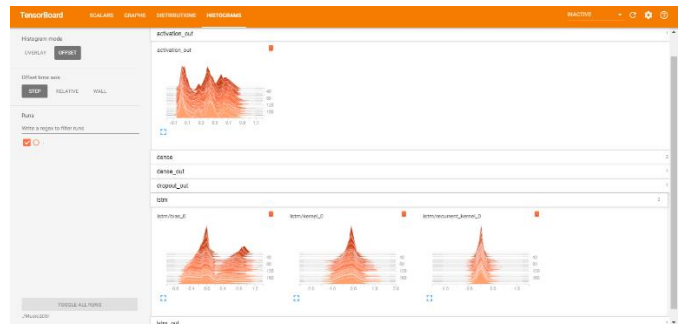

Fig. 11. Showing 3D and fancier look of the Distribution of weights and biases in the model, generated by running Tensorboard server for this model summary. Expanded view of the LSTM layer and output weights of the activation layer. Image is taken form Tensorboard.

## VI. CONCLUSION

This paper achieves the goal of designing a model which can be used to generate music and melodies automatically without any human intervention. The model is capable to recall the previous details of the dataset and generate a polyphonic music using a single layered LSTM model, proficient enough to learn harmonic and melodic note sequence from MIDI files of Pop music [25]. The model design is described with a perception of functionality and adaptability. Induction and method of training dataset for music generation is achieved through this work. Moreover, analysis of the model is also impersonated for better insights and understanding. Enhancement of model feasibility and past and present possibilities are also discussed in this paper. Future work will aim to test how well this model scales on much larger dataset, such as the Million Song Dataset [27]. We would like to observe effects on this model by adding more LSTM units and try different combinations of hyper parameters to see how well this model performs. We believe follow up research can optimize this model further with lots of computation.

## ACKNOWLEDGMENT

This research was supported by Medi-Caps University. We thank Prof. Dheeraj Rane, Medi-Caps University for comments that greatly improved manuscripts. We would also like to thank 4 "anonymous" reviewers for their so-called insights and their kind comments which helped in shaping early version of manuscript, although any errors are our own and should not tarnish the reputation of these esteemed people.